\documentclass{elsart}
\usepackage{psfig}

\newcommand{\gsim}{\;\raisebox{-0.9ex}
                   {$\textstyle\stackrel{\textstyle>}{\sim}$} \;}
\newcommand{\lsim}{\;\raisebox{-0.9ex}
                   {$\textstyle\stackrel{\textstyle<}{\sim}$} \;}
\def\bp {{ \mathbf{p} }} 
\def\ls {{\ell s}} 
\def\os {{os}} 
\def\ot {{\otimes}}

\def\half{{\textstyle{1\over 2}}}

\begin{document}

\begin{frontmatter}

\begin{flushright} 
{\bf hep-ex/0003029}\\
HEPHY--PUB 730/2000 \\ 
STPHY 35/00
\end{flushright}

\title{Multiplicity dependence of correlation functions
        in $\bar{p}p$ reactions at $\mathbf{\sqrt{s} = 630}$~GeV}
\author[HEPHY]       {B.\ Buschbeck,}
\author[STELLENBOSCH]{H.C.\ Eggers} and 
\author[ARIZONA]     {P.\ Lipa} 
\address[HEPHY]       {Institut f\"ur Hochenergiephysik,
                       Nikolsdorfergasse 18, A--1050 Vienna, Austria}
\address[STELLENBOSCH]{Dept of Physics, University of
                       Stellenbosch,7600 Stellenbosch, South Africa}
\address[ARIZONA]     {Arizona Research Labs NSMA, University of Arizona,
                       Tucson AZ 85724, USA}

\begin{abstract} 
  Discussions about Bose-Einstein correlations between decay products
  of coproduced W-bosons again raise the question about the behaviour
  of correlations if several strings are produced.  This is studied by
  the multiplicity dependence of correlation functions of particle
  pairs with like-sign and opposite-sign charge in $\bar{p}p$
  reactions at $\sqrt{s} = 630$~GeV.
\end{abstract}

\end{frontmatter}

\section{Introduction}

Recently, there has been much discussion regarding the possibility
that Bose-Einstein correlations and other interconnection effects
between decay products of different strings could affect the
measurement of the $W$ mass \cite{Al}. Since measurements are hampered
by low statistics and experimental difficulties \cite{WWD,WWA,WWL3},
the question arises whether and where effects of the superposition of
several strings can be tested independently.  Within the Dual Parton
Model \cite{DPM} for hadron-hadron reactions, the charged-particle
multiplicity N is expected to rise with the number of strings.  The
decrease in the observed correlation strength $\lambda$ of the
Bose-Einstein effect as function of multiplicity \cite{UA1,TEV} may be
explained in terms of products of different strings, where each string
symmetrizes separately~\cite{And86a,And97a}.  To test this idea
further, it is desirable to investigate quantitatively the
multiplicity dependence of like-sign particle correlations.

Improvements in experimental analysis techniques \cite{Lip96a} and
larger data samples make it possible to repeat and extend
Bose-Einstein analyses with an advanced strategy.~\cite{Mat} In this
Letter, we investigate correlation functions of particle pairs with
like-sign ($\ls$) and opposite-sign ($\os$) charges at different total
charged-particle multiplicities with the same model-independent
strategy and good statistics. The bias introduced by selecting events
of a given overall multiplicity is eliminated with the use of
``internal cumulants''~\cite{Lip96a}.

\section{Data sample and normalized density correlation functions}
\label{dcfu}

The data sample consists of 1,200,000 non-single-diffractive
$\bar{p}p$ reactions at $\sqrt{s} = 630$~GeV measured by the UA1
central detector~\cite{UA1_89a}. Only vertex-associated charged tracks
with transverse momentum $p_{T} \geq 0.15$~GeV/c, $|\eta | \leq 3$,
good measurement quality and fitted length $\geq$ 30cm have been used.
To avoid acceptance problems, we restrict the azimuthal angle to
$45^{0} \leq |\phi | \leq 135^{0}$ (``good azimuth'').  Since,
however, the multiplicity for the entire azimuthal range is the
physically relevant quantity, we select events according to their
uncorrected all-azimuth charged-particle multiplicity $N$. The
corrected multiplicity density is then estimated as twice $n$, the
charged-particle multiplicity in good azimuth: $(dN_c/d\eta) \simeq
2(dn/d\eta)$.

All quantities measured are defined in the notation of correlation
integrals~\cite{Lip92a},
\begin{equation}
\label{ptoq}
r_2 (Q)
=  {\rho_2 (Q) \over \rho_1 \ot \rho_1(Q)} 
=  
{\int_\Omega d^3 \bp_1 \, d^3 \bp_2 \, \rho_2 (\bp_1,\bp_2) \,
                      \delta \left[Q - q(\bp_1,\bp_2) \right]
    \over
    \int_\Omega d^3\bp_1\, d^3\bp_2\, \rho_1(\bp_1)\,
         \rho_1(\bp_2)\,
                      \delta \left[Q - q(\bp_1,\bp_2) \right]
}
\,,
\end{equation}
with $\bp_i$ the three-momenta, $p_i$ the corresponding four-momenta,
and $q \equiv \sqrt{-(p_1 - p_2)^2}$. As usual, $\rho_i$ denotes the
joint particle density of order $i$ which in integrated form is the
appropriate factorial moment. The integration region $\Omega$ is
identical with our experimental cuts as specified above and
specifically refers to the good-azimuth region.  All particles have
been assumed to be pions.

In $\bar{p}p$ reactions and in full phase space, the number of
positive and negative particles are equal, as are the corresponding
one-and two-particle densities $\rho_1$, $\rho_2$. Given the charged
particle density
$\rho_1(\bp)   = \rho_1^+(\bp) + \rho_1^-(\bp)$, we hence assume that 
$\rho_1^+(\bp) = \rho_1^-(\bp) = {1\over2} \rho_1(\bp)$ 
and therefore also 
$\rho_1^+ \ot \rho_1^+(Q) = {1\over 4}\rho_1\ot\rho_1(Q)$ 
etc. The like-sign and opposite-sign normalised two-particle 
densities become, respectively, 
\begin{eqnarray*} 
r_2^\ls (Q) 
&=&  {\rho_2^\ls(Q) \over \rho_1\ot\rho_1^\ls (Q) } 
=    {\rho_2^{++} (Q)  
      \over
      \rho_1^+ \ot \rho_1^+ (Q) 
     } 
=    {\rho_2^{--} (Q)  
      \over
      \rho_1^- \ot \rho_1^- (Q) 
     } 
\simeq 
     {\rho_2^{++} (Q) + \rho_2^{--}(Q)  
      \over
       \half \rho_1\ot \rho_1(Q) 
     }
\,, \\ 
r_2^{os} (Q) 
&=& {\rho_2^\os(Q) \over \rho_1\ot\rho_1^\os (Q) } 
=   {\rho_2^{+-} (Q) 
     \over
     \rho_1^+ \ot \rho_1^- (Q)  
    } 
=   {\rho_2^{-+} (Q) 
     \over
     \rho_1^- \ot \rho_1^+ (Q)  
    } 
\simeq
    {\rho_2^{+-} (Q) + \rho_2^{-+}(Q)
     \over
      \half  \rho_1\ot \rho_1(Q) 
    } \,.
\end{eqnarray*} 
Fig.\ 1 shows the normalized density correlation functions $r_2$ for pairs of
like-sign ($\ls$) charge and for opposite-sign ($\os$) charge separately. 
Restricting the total uncorrected charged-particle multiplicity $N$ in 
$|\eta | \leq 3$ to the windows 
$ 5 \leq N \leq  9$ (Fig.\ 1a) and 
$28 \leq N \leq 35$ (Fig.\ 1b), one obtains for the 
corrected particle density in the central rapidity region 
$dN_c/d\eta = 1.22 \pm 0.09$ and 
$dN_c/d\eta = 5.32 \pm 0.17$ respectively.
Both the like-sign and opposite-sign
correlation densities show a strong dependence on multiplicity.\footnote{
The usual Bose-Einstein analysis assumes that $r_2^\ls$ tends to a constant
for large $Q$, ie.\ $r_{2\ \rm{BE}}^\ls (Q \geq 1)$ = constant. It
should be clear from Figure 1 that no such constancy exists for 
limited-multiplicity windows.}

\section{Cumulants for fixed multiplicity 
and multiplicity ranges}

While the multiplicity dependence of $r_2$ is of much interest, a
quantitative analysis must both remove combinatorial background by
calculating the cumulants and also correct for the bias introduced by
working at fixed multiplicity. This bias arises because at fixed total
multiplicity $N$, standard second-order factorial cumulants, defined by
\begin{equation}
\label{eqn4} 
\kappa_2(\bp_1, \bp_2\,|N) 
= \rho_2 (\bp_1,\bp_2\,|N) - \rho_1(\bp_1\,|N) \, \rho_1(\bp_1\,|N) \,,
\end{equation} 
are nonzero even when particles are completely uncorrelated; for example,
for purely uncorrelated multinomially-distributed events,
$
\kappa_2^{\rm{mult}} (\bp_1, \bp_2\,|N) 
= 
-(1/N) \rho_1(\bp_1\,|N) \, \rho_1(\bp_2\,|N) \neq 0\,.  
$
Because these correlations result solely from the restriction of the
sample to a fixed multiplicity, they are termed ``external'' and
should be removed. The ``internal cumulants'' of Ref.~\cite{Lip96a}
\begin{eqnarray}
\label{eqn6} 
\kappa_2^I (\bp_1, \bp_2\,|N) 
&\equiv&
\kappa_2 (\bp_1,\bp_2 |N) - \kappa_2^{\rm{mult}} (\bp_1, \bp_2\,|N) 
      \nonumber \\
&=& \rho_2(\bp_1,\bp_2|N) 
    - {N(N{-}1) \over N^2}\; \rho_1(\bp_1|N) \rho_1(\bp_2|N) 
\end{eqnarray} 
correct this bias exactly: they are zero whenever the $N$ particles behave
multinomially.

The following arguments lead to a unique choice of normalization
$\rho_2^{\rm norm}$. The correctly normalized internal cumulants 
$
 K_2^I (\bp_1, \bp_2\,|N)
= \kappa_2^I (\bp_1, \bp_2\,|N) / \rho_2^{\rm norm}
$
should be independent of multiplicity $N$ whenever 
$\rho_1(\bp\,|N)$ and $\rho_2(\bp_1, \bp_2\,|N)$ depend on $N$ only 
globally i.e.\ have the same shapes (in terms of the momenta) for different
$N$. Such ``shape constancy'',
\begin{eqnarray}
\label{sa1}
\rho_1(\bp\,|N')
&=& {N' \over N}\; \rho_1(\bp|N) \,,
\\
\label{sa2}
\rho_2(\bp_1,\bp_2|N')
&=&  {N'(N'{-}1) \over N(N{-}1)}\; \rho_2(\bp_1, \bp_2\,|N) \,,
\end{eqnarray}
coupled to the requirement that
\begin{equation}
\label{sc}
K_2^I(\bp_1, \bp_2\,|N')
= K_2^I(\bp_1, \bp_2\,|N) \,,
\end{equation}
fixes the appropriate normalisation to be
\begin{equation}
\label{sc1}
\rho_2^{\rm norm}(N)
= {N(N{-}1) \over N^2}\; \rho_1(\bp_1|N) \rho_1(\bp_2|N) \,,
\end{equation}
the quantity to which $\rho_2$ defaults when $\bp_1$ and $\bp_2$
become statistically independent.  The correctly normalized internal
cumulant at fixed multiplicity consequently reads
\begin{equation}
\label{sd}
 K_2^I(\bp_1, \bp_2\,|N)
= {N^2 \over N(N{-}1)}\; {\rho_2(\bp_1, \bp_2\,|N) \over
 \rho_1(\bp_1|N) \rho_1(\bp_2|N)} \;  -  \;  1     \,.
\end{equation}
Extending these arguments to multiplicity ranges $N \in [A,B]$,
specifying them for different charge combinations and adopting the
correlation integral Eq.\ (\ref{ptoq}), we arrive at our measurement
prescription for the normalized internal cumulants for $\ls$ and $\os$
pairs,
\begin{eqnarray}
\label{se}
\overline{K_2}^{I\ls}(Q| AB) 
&=& {\overline{n}_{\ls}^2 \over \overline{{n(n{-}1)}}_{\ls}}
\; r_2^{\ls}(Q|AB) \; - \;  1  \,,
\\
\label{se1}
\overline{K_2}^{I\os}(Q |AB) 
&=& {\overline{n}_+ \, \,  \overline{n}_- \over \overline{n_+ n}_-}
\; r_2^{\os}(Q|AB) \; - \;  1    \,,
\end{eqnarray}
where $\overline{n}_{\ls}\ (= \overline{n}_+ = \overline{n}_-)$ , 
$\overline{n_+ n}_-$ and 
$\overline{{n(n{-}1)}}_{\ls}$ 
are the mean numbers of positive or negative particles, 
$\os$ pairs and ${+}{+}$ (or ${-}{-}$) pairs 
respectively in the whole interval $\Omega$
and in the multiplicity range $[A,B]$.
Eqs.\ (\ref{se}) and (\ref{se1}) are obtained \cite{tbp}
by assuming shape constancy for all averaging procedures\footnote
{The renormalization factors in front of $r_2$ were used in Ref
  \cite{WH} and elsewhere. Here, they are derived from Eqs.\ 
  (\ref{sa1}), (\ref{sa2}) and the requirement (\ref{sc}).}.

Since all quantities shown here and below are to be understood as mean
values in multiplicity ranges, we henceforth (and in Fig.\ 1) omit the
bar on the symbols.  Fig.\ 2 shows both like- and opposite-sign
internal cumulants for two selections of $dN_c/d\eta$. Three features
are apparent:
\begin{itemize}
  
\item[I.]  The importance of changing from $r_2$ to $K_2^I$ lies in
  the fact that the latter demarcate clearly the ``line of no
  correlation'' which, due to the fixed-multiplicity conditioning, is
  not equal to unity for $r_2$ in Fig. 1a.
  
\item[II.] Internal cumulants integrate to zero over the entire phase
  space; hence the positive part of $K_2^I$ at small $Q$ is
  compensated by a negative part at larger $Q$. Physically, this means
  that particles like to cluster, so that there is a surfeit of pairs
  at small $Q$ and a dearth of pairs at large $Q$ compared to the
  uncorrelated case.
  
\item[III.] The dependencies of the like- and opposite-sign cumulants
  on multiplicity are rather similar in that both decrease markedly
  with $dN_c/d\eta$. This is discussed more fully below.

\end{itemize}

\section{Multiplicity dependence of normalized cumulants}

The behaviour of $K_2^{I\, \ls}$ and $K_2^{I\, os}$ as a function of
$dN_c/d\eta$ suggests that both could have approximately the same
functional dependence $C(dN_c/d\eta)$ on multiplicity density.  Under
this hypothesis, where both depend on $dN_c/d\eta$ with the same
functional form\footnote{
For simplicity we write $N_c$ instead of $dN_c/d\eta$  here and below.},
\begin{eqnarray}
\label{eqn12} 
K_2^{I\ls}(Q | N_c) 
&=& Y^\ls (Q)  \; C(N_c, Q) \,,  \\ 
K_2^{I\os}(Q | N_c)
&=& Y^\os (Q)  \; C(N_c, Q) \,, \nonumber 
\end{eqnarray} 
the quotient of the cumulants should be independent of multiplicity,
\begin{equation}
\label{eqn13} 
{K_2^{I\ls} (Q | N_c) \over K_2^{I\os} (Q | N_c)}
\quad = \quad {Y^\ls(Q)  \over  Y^{os}(Q)} 
\quad = \quad \left( \mbox{constant in\ } N_c \right)\,.  
\end{equation} 
Figure 3a shows that (\ref{eqn13}) holds approximately. (In the
region where both cumulants are near zero, no meaningful quotients can
be formed.)

Having shown that like- and unlike-sign internal cumulants behave
approximately in the same way as functions of $dN_c/d\eta$, we now
look for an appropriate functional form for this dependence.  A first
hypothesis is that $K_2^I$ depends inversely on $N_c$,
\begin{equation} \label{eqnff} 
K_2^{I\, a}(Q\,|N_c) 
= Y^a(Q) \, C(N_c, Q) 
= Y^a(Q) \, N_c^{-1} \qquad  a = \ls, os \,.  
\end{equation}
This can be motivated theoretically by 
\begin{itemize} 

\item[a)] {\bf Resonances:} If the unnormalized cumulants
  $\kappa_2^{I\, os}$ and $\kappa_2^{I\, \ls}$ were wholly the result
  of resonance decays and if the number of resonances were
  proportional to the multiplicity $N_c$, then $\kappa_2^I \propto
  N_c$.  Assuming shape constancy $\rho_1(\bp\,|N_c) \propto N_c \,
  \rho_1(\bp)$ gives $\rho_1 \ot \rho_1 \propto N_c^2$, and hence
  after normalization, the resonance-inspired guess is
  \[ 
  K_2^I 
  = \frac{\kappa_2^{I\, {\rm res}}}  
         {\rho_1\ot\rho_1}
  \propto \frac{1}{N_c} \,.  
  \] 

\item[b)] {\bf Independent superposition} in momentum space of $\nu$
  equal strings would also lead to $K_2^I = \nu \kappa_2^{I\, {\rm
      string}} / (\rho_1\ot\rho_1) \propto (1/\nu) \propto (1/N_c)$.
  Deviations can occur if the strings are unequal or if there is no
  strong proportionality between $\nu$ and $N_c$.
\end{itemize} 
Eq.\ (\ref{eqnff}) implies that $K_2^{Ia}(Q\, |N_1)/K_2^{Ia}(Q\,|N_2)
= (N_2/N_1) =$ (constant in $Q$) for two multiplicities $N_1$ and
$N_2$ for the same $a$ ($= \ls$ or $\os$).  In Fig.\ 3b, we show the
quotient of cumulants for two multiplicities.  Surprisingly, we find
not one but two regions of approximate constancy in $Q$, one for small
$Q \lsim 0.4$~GeV, one for large $Q \gsim 2$~GeV, where only the
latter corresponds to the value $(N_2/N_1)$ expected from
(\ref{eqnff}), shown as the dotted line. At small $Q$, one must
clearly look for other functional forms for $C(N_c,Q)$. Some
phenomenological guesses are as follows.

\begin{itemize}
  
\item[c)] In the {\bf Quantum Statistical approach} of Bose-Einstein
  correlations \cite{Wei}, no multiplicity dependence is expected with
  our renormalization of $r_2$ in Eq.\ (\ref{se}) .

\item[d)] A mixture of processes a) and c) could result in a
  dependence
  \begin{equation} \label{eqn16} 
  K_2^{I\, \ls} (Q\,|N_c) 
  \approx a(Q) + {b(Q) \over N_c} \,.
  \end{equation}
  However, no comparable picture is available for the unlike-sign
  case.

\end{itemize} 
In order to test the above ideas, we plot the cumulants against
$(dN_c/d\eta)^{-1}$ as follows.  To avoid local statistical
fluctuations, the normalized cumulants $K_2^{I\, \ls} (Q)$ and
$K_2^{I\, os} (Q)$ are fitted with suitable functions in restricted
$Q$-ranges for each $dN_c/d\eta$ (not shown).  The best-fit values at
small $Q$ (0.1 GeV/c) and at large $Q$ (7 GeV/c) are plotted in Figs.\ 
4a and 4b respectively.  The $K_2^{I\, \ls}(Q)$ have also been fitted
to an exponential parametrization for $Q < 1$~GeV/c ,
\begin{equation}\label{eqn14}
K_2^{I\, \ls} (Q) 
= a + \lambda \ e^{-RQ}\,,
\end{equation} 
since the reported increase \cite{UA1,TEV} of the radius R in case of
$\ls$ (Bose-Einstein) functions could cause part of the decrease with
$dN_c/d\eta$.  The $\lambda$ values obtained are thus corrected for
effects of varying radii, but still indicate a pronounced multiplicity
dependence (crosses in Fig.\ 4a) similar to the model-independent
cumulants (filled circles)\footnote{
  The values of $R$ from the fit to Eq.\ (\ref{eqn14}) increase by
  about 30\% over the range of $dN_c/d\eta$ considered.  A stronger
  dependence of $R$ on multiplicity and strongly decreasing $\lambda$
  values have been observed when fitting Gaussian functions \cite{UA1}
  and extending the fit range to 2 GeV/c.  This indicates the
  dependence of the results on the choice of fit functions and
  regions. Here, we are emphasizing the region of small $Q$ where
  Gaussian fits fail completely \cite{UA193,Eg97}.  }.
Fig.\ 4 shows that, as in Fig.\ 3b, the $(1/N_c)$-dependence is
satisfied only for large but not for small $Q$.  The $a + b/N_c$
dependence in Fig.~4a (solid line) provides a possible, but hardly
unique, explanation.

The important region around $Q\simeq 1$~GeV, where the phase space
contributes maximally \cite{DIV} is unfortunately difficult to
investigate, because there the $K^{I\,os}_{2}$ are decreasing rapidly
with increasing $Q$ while the $K^{I\, \ell s}_{2}$ are already small
(Fig.~2).  A $1/N_c$-dependence (due to resonances such as $\rho^{0}$)
in this dominant region around 1 GeV/c could presumably cause the
large-$Q$ region to follow suit via missing pairs.

\section{Summary}

The multiplicity dependence of like-sign and opposite-sign two-body
correlation functions have been studied with the same
model-independent strategy. The bias introduced by selecting events of
a given overall multiplicity is eliminated by measuring internal
cumulants.  We observe that
\begin{itemize} 
  
\item[--] the like-sign and opposite-sign cumulants have very similar
  multiplicity dependence,
  
\item[--] there exist two regions, one at small $Q$, where the
  multiplicity dependence of both is weaker than $1/N_c$, and one at
  large $Q\ (\gsim 2$~GeV/c), where the cumulants are negative and
  follow roughly an $1/N_c$ law,
  
\item[--] a third region arround $Q$ = 1~GeV/c shows small and rapidly
  changing cumulants.
\end{itemize}

The decrease of $\ls$ functions at small $Q$ with increasing
multiplicity favours an interpretation in terms of a suppression of
Bose-Einstein correlations between products of different strings.  In
the Dual Parton Model approach \cite{DPM}, multiparton collisions
corresponding to multipomeron exchange are expected to contribute to
the inelastic cross section. The observed energy dependence of
multiplicity distributions supports this view \cite{Ma99}.  Up to 2-3
pomeron exchanges might occur at our highest multiplicities.  This
could explain quantitatively the corresponding suppression of $\ls$
(Bose-Einstein) functions in Fig.\ 4a.  But this interpretation is not
unique. Adopting e.g.\ the ``core-halo picture'' \cite{CSO}, one could
explain the observations by assuming that long-lived resonances (such
as the $\omega$ and $\eta$) are produced more frequently at larger
$N_c$ thereby increase the relative strength of the halo \cite{Bi00}
and hence do not contribute to Bose-Einstein correlations.  Resonance
decays (Regge terms) should arguably also contribute to
$\ls$-functions \cite{Ber77a}.  One could therefore try to explain
alternatively the decrease of $\ls$ functions with $N_c$ by a mixture
of Bose-Einstein correlations (assumed to be constant in $N_c$) with
resonance production.

All this leaves unexplained, however, the similar behaviour of $\os$
and $\ls$ functions.  Theoretical work \cite{DPM,And86a,And97a,Suz99a}
is challenged by the above experimental results. Together with the
results of other reactions, they represent a new piece of information
\cite{IS99} in the colourful puzzle of multiparticle production.

\section*{Acknowledgements} 
We thank B.\ Andersson, A.\ Bia\l as and W.\ Kittel for useful
discussions, and gratefully acknowledge the technical support of G.\ 
Walzel.  HCE thanks the Institute for High Energy Physics in Vienna
for kind hospitality. This work was funded in part by the South
African National Research Foundation.


\newpage

\begin{figure} 
\psfig{figure=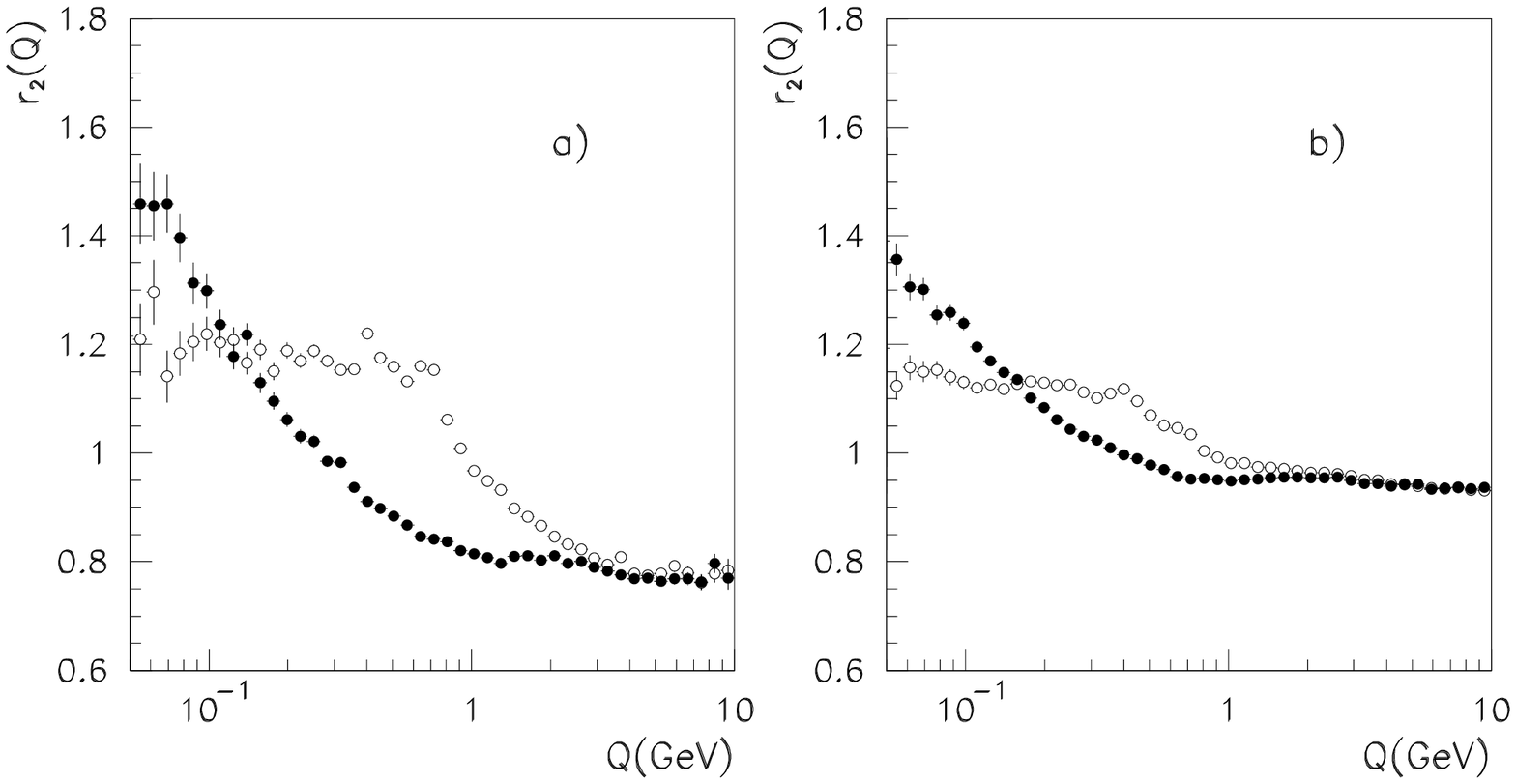,height=80mm}
\caption{
  Normalised moments
  $r_2^\ls$ versus $Q$ for like-sign pairs (filled circles) and 
  $r_2^\os$ versus $Q$ for opposite-sign pairs (open cirles) at 
  a) $dN_c/d\eta = 1.2$ and 
  b) $dN_c/d\eta = 5.3$.} 
\end{figure}

\begin{figure} 
\psfig{figure=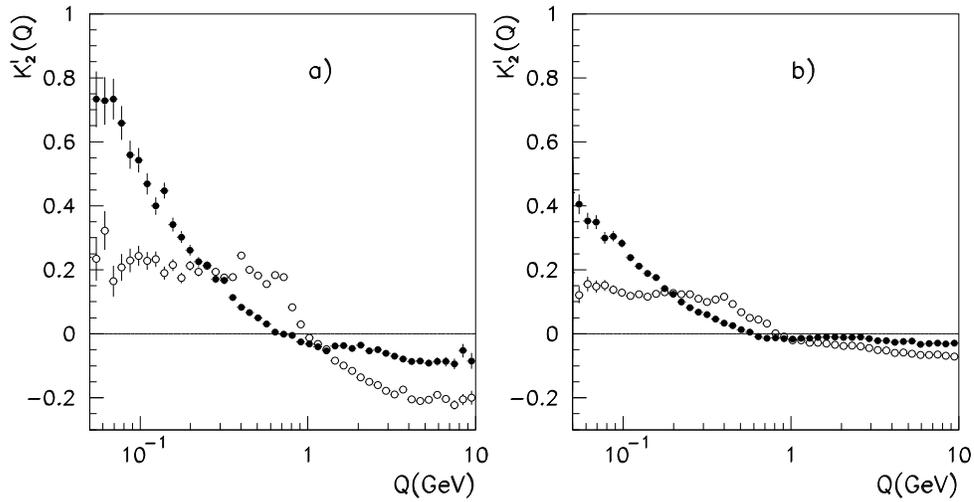,height=80mm}
\caption{
  Internal cumulants Eqs.~(\ref{se})--(\ref{se1}) for $\ls$ pairs
  (filled circles) and $\os$ pairs (open circles) for multiplicity
  densities
  a) $dN_c/d\eta = 1.2$ and 
  b) $dN_c/d\eta = 5.3$.}
\end{figure}

\begin{figure}
\psfig{figure=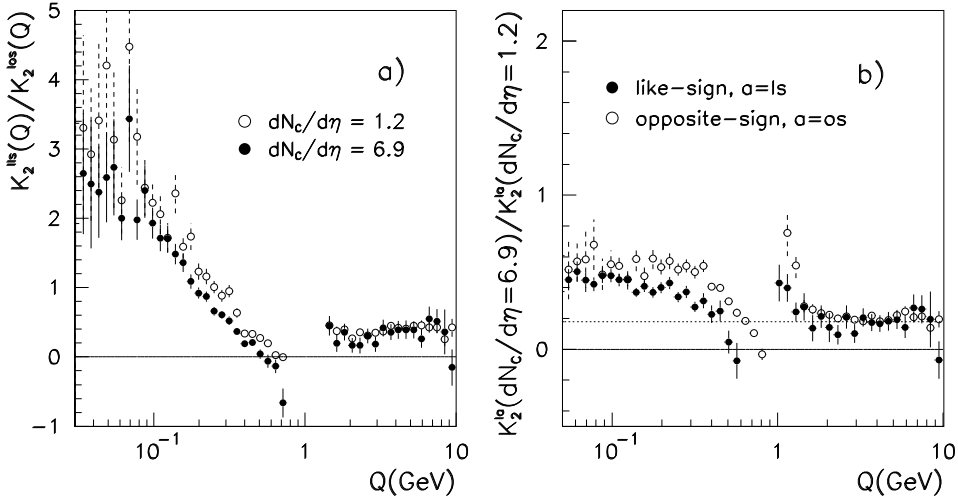,height=80mm}
\caption{
  a) The ratio Eq.\ (\ref{eqn13}), shown for two selections of $dN_c/d\eta$.  
  b) Ratio of two like-sign and two opposite-sign cumulants at different
  multiplicities $dN_c/d\eta$. The region around $Q=1$ GeV has been
  omitted because both numerators and denominators are near zero.}
\end{figure}

\begin{figure} 
\psfig{figure=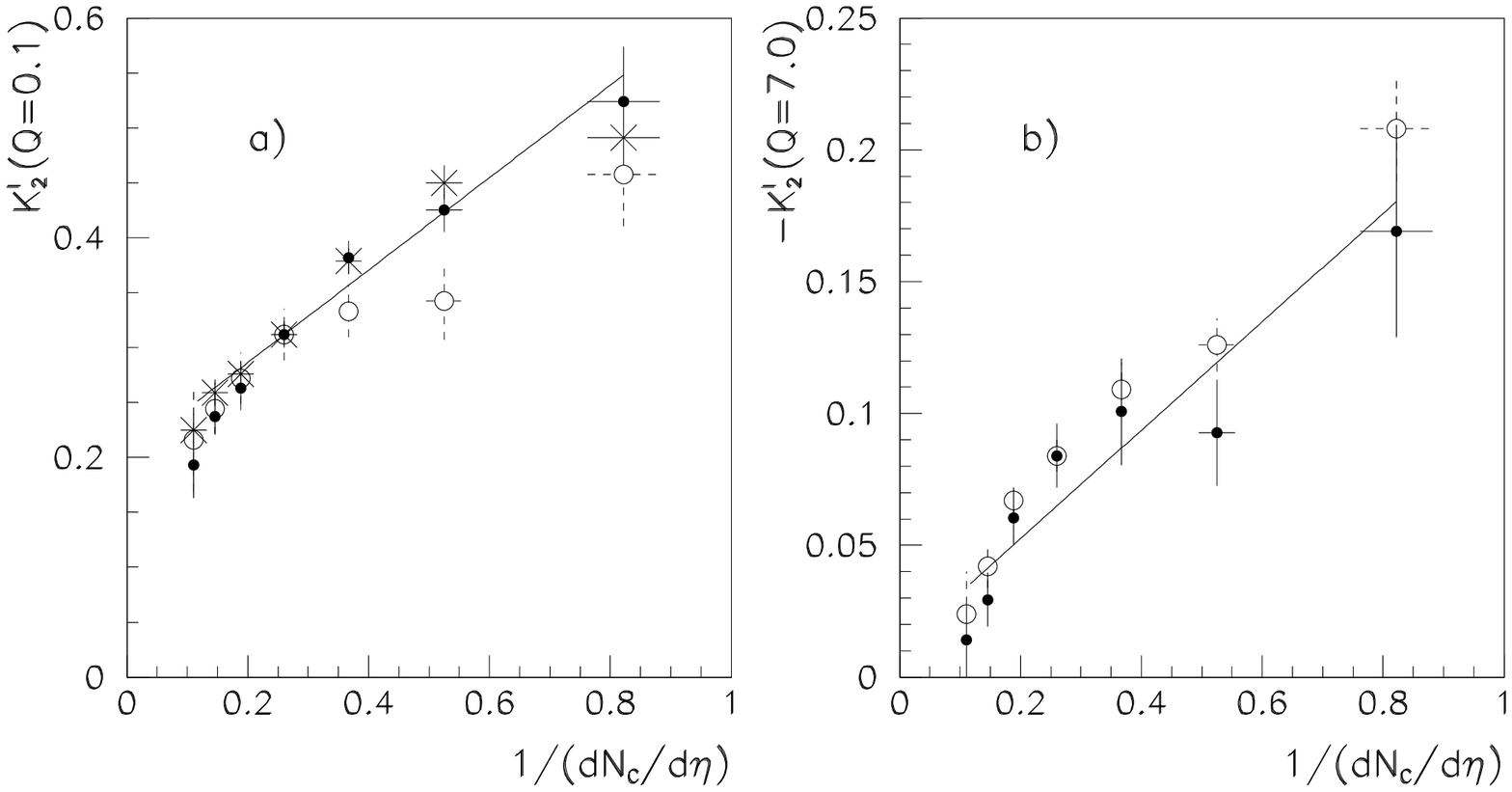,height=80mm}
\caption{
  a) Multiplicity dependence of $K_2^{I\ls}$ (filled circles) and
  $K_2^{I\os}$ (open circles), both at $Q=0.1$ GeV, plus $\lambda$
  values (crosses) of exponential fit (\ref{eqn14}).  
    b) as in a) but for the large $Q=7$~GeV-region.  Note from Fig.\ 2
  that the cumulants are negative at large Q.  For better comparison
  of the respective dependencies on $dN_c/d\eta$, the absolute values
  have been scaled by constant factors.  Solid lines are best fits to
  the $\lambda$'s using Eq.\ (\ref{eqn14}).  }
\end{figure}


\begin{thebibliography}{99}

\bibitem{Al} A.\ de Angelis in:
             Proc. 26th Internat.\ Symposium on Multiparticle 
             Dynamics, Frascati,Italy,
             edited by G.\ Capon, V.A.\ Khoze, G.\ Panchieri and S.\ Sansoni,
             {\it Nucl.\ Phys.}\ B (Proc.Suppl.), 71 (1999).
              

\bibitem{WWD} W.\ Kittel, invited talk,
              F.\ Martin, invited talk,
                at 34th Rencontre de Moriond, ``QCD and 
                High Energy Hadronic Interactions'', Les Arcs (France),
                March 20--27, 1999. 

\bibitem{WWA} ALEPH Collaboration, R.\ Barate et al.,
              CERN-EP-99-173, 1999. 

\bibitem{WWL3} L3 Collaboration, submitted to the International  
               Europhysics Conference on High Energy Physics 1999,
               Tampere, Finland; L3 Note 2405.


\bibitem{DPM} A.\ Capella U.\ Sukhatme, C.-I.\ Tan and J.\ Tran Thanh Van, 
            {\it Phys.\ Rep.\/}  {\bf 236}, 225-329 (1994).

\bibitem{UA1}UA1 Collaboration, C.\ Albajar et al.,
            {\it Phys.\ Lett.\/} B{\bf 226}, 410 (1989).

\bibitem{TEV} T.\ Alexopoulos et al., 
            {\it Phys.\ Rev.\/} D{\bf 48}, 1931 (1993).

\bibitem{And86a}B.\ Andersson and W.\ Hofmann, 
            {\it Phys.\ Lett.\/} B{\bf 169}, 364 (1986).

\bibitem{And97a} B.\ Andersson and M.\ Ringn\'er, 
            {\it Nucl.\ Phys.}\ B{\bf 513}, 627 (1997);\\
            J.\ H\"akkinen and M. Ringn\'er, 
            {\it Eur.\ Phys.\ J.}\ C{\bf 5}, 275 (1998).

\bibitem{Lip96a} P.\ Lipa, H.C.\ Eggers, and B.\ Buschbeck, 
            {\it Phys.\ Rev.\/} D{\bf 53}, R4711 (1996).

\bibitem{Mat}B.\ Buschbeck, H.C.\ Eggers and P.\ Lipa in:
            {\it Correlations and Fluctuations}, Proc.\
            8th International Workshop on Multiparticle Production, 
            M\'atrah\'aza, Hungary, 
            edited by T. Cs\"org\H o, S. Hegyi, G. Jancs\'o and R.C. Hwa,
            World Scientific (1999), pp.\ 28--36.

\bibitem{UA1_89a}UA1 Collaboration, C.\ Albajar et al.,
            {\it Z.\ Phys.\/} C{\bf 44}, 15 (1989);\\
            M.\ Calvetti et al., 
            {\it IEEE Trans. Nucl. Science NS--30}, 71 (1983).

\bibitem{Lip92a} P.\ Lipa et al., 
            {\it Phys.\ Lett.\/} B{\bf 285}, 300 (1992);\\
            H.C.\ Eggers et al., 
            {\it Phys.\ Lett.\/} B{\bf 301}, 298 (1993);\\
            H.C.\ Eggers et al., 
            {\it Phys.\ Rev.\/} D{\bf 48}, 2040 (1993).

\bibitem{tbp}B.\ Buschbeck, H.C.\ Eggers and P.\ Lipa, to be published.

\bibitem{WH}A.\ Gyulassy, S.K.\ Kauffmann and L.W.\ Wilson,
            {\it Phys.\ Rev.\/} C{\bf 20}, 2267 (1979);
             A.\ Wiedemann and U.\ Heinz, CERN-TH/99-15,
             submitted to Phys.Rep.

\bibitem{Wei} G.N.\ Fowler et al.,
            {\it Phys.\ Lett.\/} B{\bf 253}, 421 (1991).


\bibitem{UA193}UA1 Collaboration, N.\ Neumeister et al.,
            {\it Z.\ Phys.\/} C{\bf 60}, 633 (1993).

\bibitem{Eg97}  H.C.\ Eggers, P.\ Lipa and B.\ Buschbeck, 
            {\it Phys.\ Rev. Lett.\/} {\bf 79}, 197 (1997).

\bibitem{DIV} B.\ Buschbeck, P.\ Lipa and F.\ Mandl in:
              Proc.\ Nato Advanced Research Workshop on 
              Hot Hadronic Matter: Theory and Experiment,
              Divonne, France 1994; Edited by J.\ Letessier, H.H.\ Gutbrod
              and J.\ Rafelski, Plenum Press (New York) 1995.

\bibitem{Ma99} S.\ Matignyan and W.D.\ Walker ,
            {\it Phys.\ Rev.\/} D{\bf 59}, 034022 (1999).

\bibitem{CSO} T.\ Cs\"org\H o,
            {\it Phys.\ Lett.\/} B{\bf 409}, 11 (1997).

\bibitem{Bi00} A.\ Bia\l as, private comunication

\bibitem{Ber77a} E.L.\ Berger et al.,
             {\it Phys.\ Rev.\/} D{\bf 15}, 206 (1977)

\bibitem{Suz99a}N.\ Suzuki and M.\ Biyajima,
            {\it Phys.\ Rev.} C{\bf 60}, 034903 (1999).

\bibitem{IS99} B.\ Buschbeck and H.C.\ Eggers in:
             Proc.\ 29th Internat.\ Symposium on Multiparticle 
             Dynamics, August 1999, Brown Univ., Providence RI, USA,
             to be published. 


\end{thebibliography}
\end{document}